\documentclass[aps,pra,amsmath,amssymb,reprint]{revtex4-1}
\usepackage{graphicx}
\usepackage{dcolumn}
\usepackage{bm}
\usepackage{epstopdf}

\draft 

\begin{document}





\title{Measurement of the Gravity-Field Curvature by Atom Interferometry} 








\author{G. Rosi$^1$}
\author{L. Cacciapuoti$^2$}
\author{F. Sorrentino$^1$}
\altaffiliation[Present address: ]{INFN Sezione di Genova, Via Dodecaneso 33, I-16146 Genova, Italy}
\author{M. Menchetti$^1$}
\altaffiliation[Present address: ]{School of Physics and Astronomy, University of Birmingham, Edgbaston, Birmingham, B15 2TT, United Kingdom}
\author{M. Prevedelli$^3$}
\author{G. M. Tino$^1$}
\email[]{tino@fi.infn.it}
\affiliation{$^1$Dipartimento di Fisica e Astronomia and LENS, Universit\`a di Firenze, INFN Sezione di Firenze, via Sansone 1, I-50019 Sesto Fiorentino, Firenze, Italy \\ $^2$European Space Agency, Keplerlaan 1, 2200 AG Noordwijk, Netherlands \\ $^3$Dipartimento di Fisica e Astronomia, Universit\`a di Bologna, Via Berti-Pichat 6/2, I-40126 Bologna, Italy}








\date{\today}

\begin{abstract}
We present the first direct measurement of the gravity-field curvature based on three conjugated atom interferometers. Three atomic clouds launched in the vertical direction are simultaneously interrogated by the same atom interferometry sequence and used to probe the gravity field at three equally spaced positions. The vertical component of the gravity-field curvature generated by nearby source masses is measured from the difference between adjacent gravity gradient values. Curvature measurements are of interest in geodesy studies and for the validation of gravitational models of the surrounding environment. The possibility of using such a scheme for a new determination of the Newtonian constant of gravity is also discussed.
\end{abstract}


\maketitle 





In the last two decades, atom interferometry \cite{Varenna2014} has profoundly changed precision inertial sensing, leading to major advances in metrology and fundamental and applied physics. The outstanding stability and accuracy levels \cite{Peters1999, Sorrentino2014} combined with the possibility of easily implementing new measurement schemes \cite{Muller2009, Debs2011, Poli2011, Kovachy2012} are the main reasons for the rapid progress of these instruments. Matter-wave interferometry has been successfully used to measure local gravity \cite{Peters2001}, gravity gradient \cite{McGuirk2002, Sorrentino2012,Duan2014}, the Sagnac effect \cite{Gauguet2009}, the Newtonian gravitational constant \cite{Bertoldi2006, Fixler2007, Lamporesi2008, Rosi2014}, the fine structure constant \cite{Bouchendira2011}, and for tests of general relativity \cite{Schlippert2014, Tarallo2014}. Accelerometers based on atom interferometry have been developed for many practical applications including geodesy, geophysics, engineering prospecting, and inertial navigation \cite{DeAngelis2009, Bresson2006, Geiger2011}. Instruments for space-based research are being conceived for different applications ranging from weak equivalence principle tests and gravitational-wave detection to geodesy \cite{Tino2013, Graham2013}.

One of the most attractive features of atom interferometry sensors is the ability to perform differential acceleration measurements by simultaneously interrogating two separated atomic clouds with high rejection of common-mode vibration noise, as demonstrated in gravity gradiometry applications \cite{McGuirk2002,Sorrentino2014}. In principle, such a scheme can be extended to an arbitrary number of samples, thus, providing a measurement of higher-order spatial derivatives of the gravity field.
Geophysical models of the Earth's interior rely on the inversion of gravity and gravity gradient data collected at or above the surface \cite{Hofmann2005}. The solution to this problem, which is, in general, not unique, leads to images of the subsurface mass distribution over different scale lengths \cite{Rummel2011}. In this context, the simultaneous determination of gravity acceleration and its derivatives improves the inversion procedure by introducing additional constraints for the valid solutions. Gravity gradient surveys are already used to detect short-wavelength density anomalies or in situations where the vibration noise seriously limits absolute gravity measurements. The second derivative of the gravity field can vary by several orders of magnitude when measured across shallow density anomalies, promising high spatial resolutions and sharp signals for their localization \cite{Butler1984}. Simultaneous \textit{in situ} measurements of the gravity acceleration and its derivatives can also be used for remote sensing to estimate the evolution of the gravitational field along the direction of the local plumb line. Such a method could find interesting applications in regional height systems to measure differences in the gravitational potential with respect to a reference station, e.g., located on the geoid \cite{Rummel2002}. Indeed, in the presence of shallow density anomalies, the knowledge of both the gravity gradient and the curvature can provide centimeter-level resolution ($\sim 0.1\ \textrm{m}^2/\textrm{s}^2$) in the measurement of differential geopotential heights by integrating the gravity field over baselines of several hundreds of meters. The simultaneous measurement of gravity gradient and higher-order derivatives would also help with correcting for Newtonian noise in future gravitational-wave detectors \cite{Saulson1984}.

In this Letter, we report for the first time the direct measurement of the gravity-field curvature generated by nearby source masses, as suggested in Ref. \cite{McGuirk2002}. Our atom interferometer, which simultaneously probes three freely falling samples of $^{87}$Rb, is able to perform measurements of gravity, gravity gradient, and curvature along the vertical direction at the same time, opening new perspectives for geodesy studies and Earth monitoring applications. Using this scheme, we also demonstrate a new method to measure the Newtonian constant of gravity.

The details of the experimental apparatus can be found in Refs. \cite{Rosi2014, Sorrentino2010}. In the following, a description of the measurement sequence and data analysis will be provided, with particular emphasis on the new features introduced by the third atomic sample and the gravity curvature determination.

\begin{figure}[t!]
\includegraphics[width=0.49\textwidth]{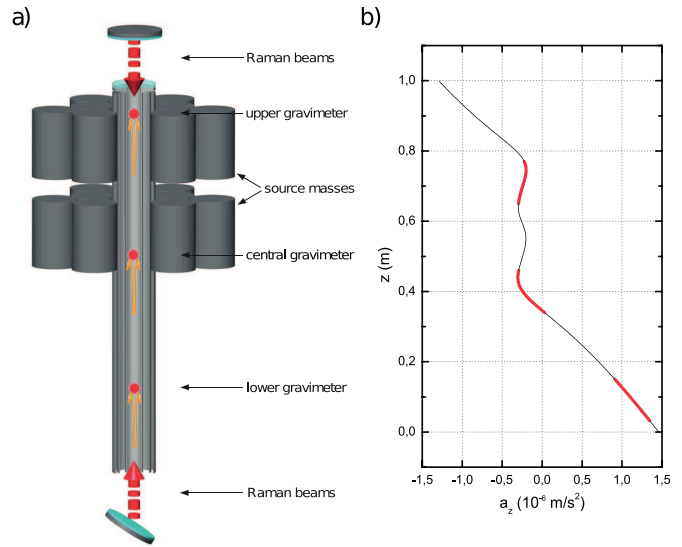}
\caption{\label{setup} (color online) a) Scheme of the experiment. $^{87}$Rb atoms are trapped and cooled in a MOT. Three atomic clouds are launched in rapid sequence along the vertical direction with a moving optical molasses. Near the apogees of the atomic trajectories, a measurement of the vertical acceleration sensed by the three clouds is performed by Raman interferometry. External source masses are positioned in order to maximize the average gravity curvature at the three clouds' positions. b) Gravitational acceleration along the symmetry axis ($a_z$) produced by the source masses and the Earth's gravity gradient; a constant value accounting for the Earth's gravitational acceleration was subtracted. The spatial regions of the three atom interferometers are indicated by the thick red lines.}
\end{figure}

A magneto-optical trap (MOT) with beams oriented in a 1-1-1 configuration collects $^{87}$Rb atoms and launches them vertically at a temperature of about 4 $\mu$K. A high-flux source based on a 2D MOT provides large atom numbers ($\sim10^9$) in short loading times ($\sim40$ ms). Larger atom numbers could be obtained by using the juggling technique \cite{Sorrentino2010}; however, only a direct launch can be readily implemented in our measurement cycle and extended to three or more samples. We launch three atomic clouds along the vertical direction separated by $\sim30$ cm, which reach the apogees of their atomic trajectories simultaneously. A series of velocity selection and blow-away pulses prepares the samples in the magnetically insensitive $|F=2,m_F=0\rangle$ sublevel of the $^{87}$Rb ground state. The Mach-Zehnder interferometer simultaneously addresses the three clouds with a $\pi/2-\pi-\pi/2$ sequence of vertical velocity-selective Raman pulses \cite{Kasevich1991}. The Raman lasers, with effective wave vector $k_{\textrm{eff}}\simeq16\times10^6$ m$^{-1}$, are resonant with the 6.8 GHz two-photon transition $|F=2,m_F=0\rangle\rightarrow|F=1,m_F=0\rangle$ of the $^{87}$Rb ground state and have a 2 GHz red detuning with respect to the $5^2\textrm{S}_{1/2}|F=2\rangle\rightarrow5^2\textrm{P}_{3/2}|F=3\rangle$ transition to the excited state. The sequence has a duration of $2T=320$ ms. The $\pi$ pulse lasts 24 $\mu$s and occurs 5 ms after the atomic clouds have reached their apogees. The interference fringes are obtained by measuring the normalized population in one of the two hyperfine levels of the $^{87}$Rb ground state. We use a set of high-density source masses, for a total of 516 kg, to enhance the gravity-field curvature sensed by the three atomic samples. The source masses are composed of 24 tungsten alloy (INERMET IT180) cylinders \cite{Lamporesi2007}. They are positioned on two titanium platforms and distributed in hexagonal symmetry around the axis of the interferometer tube. The vertical position of the platforms is accurately controlled by precision screws synchronously driven by stepper motors and measured by an optical readout system. The experimental setup is shown in Fig. \ref{setup}, together with the axial acceleration profile due to the source masses and the Earth's gravity.

Atomic gravity gradiometers use the same Raman lasers to simultaneously probe two spatially separated atom clouds on the same interferometric sequence. In this configuration, vibration noise that couples into the phase of the Raman lasers is seen as common mode and can be efficiently rejected. As a consequence, when the normalized atomic populations $(x,y)$ simultaneously measured at the two spatially separated interferometers are plotted in 2D space, an ellipse is obtained. Common-mode phase noise affecting the fringes of the two atom interferometers distributes the experimental points around the ellipse, whose shape carries information on the gravity gradient between the two clouds \cite{Foster2002}. We extend this idea by introducing a third atom interferometer. In this case, the normalized atomic populations $(x,y,z)$ measured at the output ports of the three atom interferometers are distributed around a Lissajous ellipse lying in 3D space.

The ellipse best fitting the data points in 3D space can be expressed with the parametric equations
\begin{equation}
\label{eq1}
\begin{cases} x(\theta)=A+B\sin\theta\,,\\ y(\theta)=C+D\sin(\theta+\varphi_1)\,,\\ z(\theta)=E+F\sin(\theta+\varphi_1+\varphi_2)\,.\end{cases}
\end{equation}
Here $A$, $B$, $C$, $D$, $E$, and $F$ represent the amplitude and offset of the fringes of the three atom interferometers, $\theta$ is the phase angle parameter, which varies randomly due to common-mode vibration noise, and $\varphi_1$ and $\varphi_2$ are the phase angles proportional to the differential accelerations between adjacent interferometers. Fitting an ellipse to points in 3D space can be recast as a 2D problem \cite{Jiang2005}. Given a set of $n$ data points $(x_i, y_i, z_i)$, the $\chi^2$ function can be written as
\begin{equation}
\label{eq2}
\chi^2\propto\sum\limits_{i=1}^n [e^{2}_{i,\textrm{ellipse}}(x_i,y_i,z_i)+h_{i}^{2}]\,,
\end{equation}
where $e_{i,\textrm{ellipse}}$ denotes the Euclidean 2D distance between the ellipse and the projection of the point on the plane of the ellipse, and $h_i$ is the 3D point-plane distance. We assume equal uncertainties on all experimental points. The $\chi^2$ function is then evaluated and minimized with respect to the eight parameters of Eq. \ref{eq1}. This approach can be easily generalized to $N$ interferometers $(x_1,...,x_N)$ for the measurement of higher-order derivatives of the gravity field along the vertical axis.

\begin{figure}[t]
\includegraphics[width=0.45\textwidth]{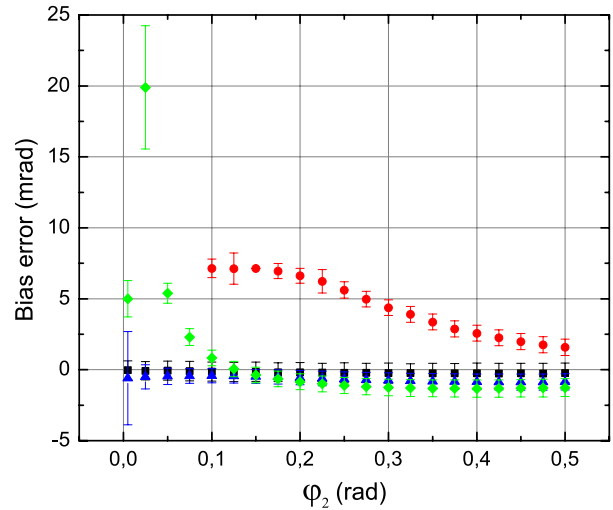}%
\caption{\label{dati_sintetici}(color online) Bias error in the differential phase from the 2D elliptical fit (red circles), the new 3D fit routine (black squares), and the Bayesian method for different $\varphi_2$ angle values; $\varphi_1$ is kept fixed at $\pi/2$. Synthetic data are generated with a Gaussian detection noise ($\sigma_d=0.01$) and a fringes contrast of $0.3$. In the Bayesian analysis, we feed the algorithm both with the exact Gaussian noise as used in the simulation (blue triangles) and with the value obtained after introducing a 10\% error on $\sigma_d$ (green rhombi). For $\varphi_2<0.1$ rad, the $\chi^2$ numerical minimization fails in the 2D fit. The knowledge of the noise model affecting the data becomes critical in the Bayesian analysis when the phase angle approaches zero.}
\end{figure}

It is worth pointing out that adding an extra dimension (i.e., a third atom interferometer) opens the possibility of accurately measuring small gradiometric phase shifts introducing negligible bias on the fit results. In a two-cloud configuration and in the presence of a small gravity gradient, the two output fringes are almost in phase and the ellipse degenerates to a line. On the other hand, in the presence of a third cloud, even if $\varphi_2\sim0$, $\varphi_1$ can be made quite large, e.g., by pulsing a magnetic field at the location of the third interferometer. In this case, $|\varphi_2-\varphi_1|\gg0$ and, even in presence of noisy data, $\varphi_2$ can be reliably extracted from the fit in 3D space. Figure \ref{dati_sintetici} compares the bias errors introduced by three different methods: a least-square fit in 2D, a least-square fit in 3D space, and a Bayesian analysis, all as a function of $\varphi_2$, when $\varphi_1$ is kept at $\pi/2$. Simulated data are affected by significant Gaussian noise at detection ($\sigma_d = 0.01$) and present a fringe contrast of $0.3$. The plot clearly shows how the third interferometer becomes instrumental for precision gravity gradiometry. The bias errors introduced by the 2D fit are non-negligible. In addition, for $\phi_2<0.1$, the 2D fit routine fails to converge. The Bayesian analysis performs better than the 2D fit, but it introduces significant biases at small phase angles when the \textit{a priori} knowledge of the noise affecting the data varies by 10\%. On the other hand, the 3D fit is very robust and, in contrast with the Bayesian method, does not require any \textit{a priori} knowledge of the noise on the experimental data.

\begin{figure}[top!]
\includegraphics[width=0.55\textwidth]{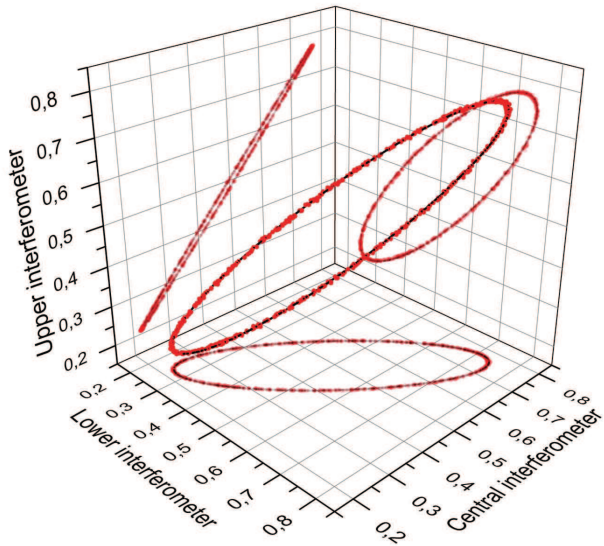}%
\caption{\label{3Dplot} (color online) Typical three-dimensional Lissajous figure obtained by plotting the output signal of the upper atom interferometer as a function of the lower and central one (red circles) and ellipse in 3D best fitting the data (black line). Orthogonal projections on the three Cartesian planes are also shown.}%
\end{figure}

Our setup has been used to perform a direct measurement of the gravity-field curvature generated by the source masses. One of the most critical aspects of the measurement is the presence of spurious and nonhomogeneous magnetic fields in the interferometer region. Because of the spatial separation between the three atomic clouds, primarily imposed by the MOT loading time, the gravity curvature measurement is averaged over a total distance of about 60 cm. In this configuration, the lower and the upper atomic samples are close to the edges of the $\mu$-metal shield surrounding the vertical tube, where the passive attenuation of external magnetic fields is lower and the internal bias field is less homogeneous. To reduce this source of systematic errors, the sign of the effective wave vector $k_{\textrm{eff}}$ of the Raman lasers is periodically reversed during data acquisition \cite{Louchet-Chauvet2011}. This is achieved by selecting a different Raman counterpropagating beam pair by properly adjusting the frequency detuning to compensate for the Doppler shift induced by the atomic motion in the gravity field. In this way, phase shifts that do not depend upon the effective wave vector, e.g., second-order Zeeman shifts or ac Stark shifts, are rejected when taking the difference between measurements performed with opposite $k_{\textrm{eff}}$. Submillimetric vertical overlap of the interferometer's arms has been achieved by properly adjusting the Raman frequency ramp and the timing of the velocity selection. Their transverse overlap is ensured by using the same launch sequence. The horizontal velocity spread due to the finite transverse atomic temperature is expected to introduce noise and systematic shifts on the ellipse phase angle via the Coriolis acceleration. Because of the double differential nature of the gravity-field curvature measurement, the effect of Coriolis accelerations depends on the difference between the relative initial velocities of the atomic clouds at the two adjacent gravity gradiometers. To further reduce this effect, the mirror retroreflecting the Raman laser beams is rotated to compensate for the Earth's rotation \cite{Hogan2007,Lan2012}. Two data sets of 720 points (2.5 s of measurement time per point), one for each of the two $k_{\textrm{eff}}$ opposite directions ($\uparrow$ and $\downarrow$ ), have been collected and analyzed. Figure \ref{3Dplot} shows a typical plot of the data points measured at the three conjugated atom interferometers, together with the ellipse in 3D space best fitting the data. The values for $\varphi_1$ and $\varphi_2$ are given by
\begin{equation}
\varphi_1=(\varphi_{1,\uparrow}-\varphi_{1,\downarrow})/2\,\,,\,\,\,\,\varphi_2=(\varphi_{2,\uparrow}-\varphi_{2,\downarrow})/2\,\,.
\end{equation}
From the measurement of the clouds' separation, $d=(0.3098\pm0.0002)$ m, it is possible to evaluate the average gravity gradients $\gamma_{1,2}=\varphi_{1,2}/(dk_{\textrm{eff}}T^2)$, obtaining $\gamma_1=(-4.112\pm0.008)\times10^{-6}$ s$^{-2}$ and $\gamma_2=(0.223\pm0.003)\times10^{-6}$ s$^{-2}$, and, thus, the average gravity curvature $\zeta=(1.399\pm0.003)\times10^{-5}$ s$^{-2}$m$^{-1}$. This measurement is consistent with the value $\zeta_{\textrm{sim}}=1.397\times10^{-5}$ s$^{-2}$m$^{-1}$ obtained from our Monte Carlo model \cite{Prevedelli2014}, which accounts for the source masses and additional contributions in the immediate vicinity of the atomic clouds.

The measurement of the second derivative of the gravity acceleration is also an interesting tool for determining the Newtonian gravitational constant $G$, as proposed in Ref. \cite{Rothleitner2014}. The method consists of performing two simultaneous gravity gradient measurements in the presence of heavy source masses. The Earth's gravity gradient contribution is rejected when calculating the difference between the two measurements without any need for modulating the position of the masses. In this way, systematic effects introduced by deformations and tilts of the structure holding the masses can be removed. For such an experiment, it becomes important to optimize the distribution of the source masses to generate three quasistationary regions to host the conjugated atom interferometers, thus, reducing the systematics arising from the positioning errors of the atomic clouds. Even if not specifically designed for this purpose, we used our apparatus to perform a proof-of-principle experiment. With the source masses and the atomic clouds positioned as in Fig. 1, we measured $\Phi_{\textrm{meas}}=\varphi_2-\varphi_1$ and compared it with $\Phi_{\textrm{sim}}$ obtained from our single-particle Monte Carlo simulation. We obtained $\Phi_{\textrm{meas}}=(0.5533\pm0.0006)$ rad, which is in good agreement with $\Phi_{\textrm{sim}}=0.5528$ rad. The short-term sensitivity of $3.8\times10^{-2}G$ at 1 s is comparable with the one obtained in Refs. \cite{Sorrentino2014, Rosi2014, Prevedelli2014} by alternating the source masses position. An extensive evaluation of the systematic error sources that are affecting the measurement is beyond the scope of this work.

In conclusion, by using three simultaneous atom interferometers, we have measured for the first time the component of the gravity curvature produced by nearby source masses along one axis. The new analysis method based on an elliptical fit in 3D space has proven to be very robust with respect to amplitude noise and immune from noise-induced systematic shifts. The scheme has also been used to perform a proof-of-principle measurement of the Newtonian gravitational constant based on two simultaneous gravity gradient measurements. Sensitivity and long-term stability of the $G$ measurement are comparable with our previous work, opening the possibility for further improvements after optimization of the distribution of the masses and the position of the atomic clouds. This method can be extended to multiple interferometers with small spatial separation ($\sim5-10$ cm) in order to reconstruct acceleration profiles with high resolution and measure higher-order derivatives of the gravitational acceleration.









%




%











%







\begin{acknowledgments}
This work was supported by INFN (MAGIA experiment), EC (FINAQS STREP/NEST project Contract No. 012986) and ESA (SAI project Contract No. 20578/07/NL/VJ). The authors acknowledge M. De Angelis, R. del Aguila, J. Flury, C. Rothleitner, and G. Saccorotti for a critical reading of the manuscript.
\end{acknowledgments}

%


\end{document}